\documentclass[reprint,superscriptaddress,amsmath,amssymb,aps,pra,floatfix]{revtex4-1}
\usepackage{graphicx}
\usepackage{array}
\usepackage{dcolumn}
\usepackage{bm}
\usepackage{listings}
\usepackage{scalefnt}
\usepackage{xcolor}
\usepackage{booktabs}
\usepackage{multirow}
\usepackage{tensor}
\usepackage{hyperref}
\usepackage{mathtools}
\hypersetup{
    colorlinks,
    linkcolor={blue!80!black},
    citecolor={blue!80!black},
    urlcolor={blue!80!black}
}
\usepackage{color}
\newcommand{\minus}{\scalebox{0.75}[1.0]{\( - \)}}
\usepackage{physics}
\usepackage[pagewise]{lineno}

\begin{document}

\title{Experimental linear optical computing of the matrix permanent}

\author{Yosep Kim}
 \affiliation{Department of Physics, Pohang University of Science and Technology (POSTECH), Pohang 37673, Korea}

 \author{Kang-Hee Hong}
 \affiliation{Department of Physics, Pohang University of Science and Technology (POSTECH), Pohang 37673, Korea}

 \author{Joonsuk Huh}
 \email{joonsukhuh@gmail.com}
 \affiliation{Department of Chemistry, Sungkyunkwan University, Suwon 16419, Korea}

\author{Yoon-Ho Kim}
\email{yoonho72@gmail.com}
 \affiliation{Department of Physics, Pohang University of Science and Technology (POSTECH), Pohang 37673, Korea}

\date{\today}
             
\begin{abstract}
Linear optical computing (LOC) with thermal light has recently gained attention because the problem is connected to the permanent of a Hermitian positive semidefinite matrix (HPSM), which is of importance in the computational complexity theory. Despite the several theoretical analyses on the computational structure of an HPSM in connection to LOC, the experimental demonstration and the computational complexity analysis via the linear optical system have not been performed yet. We present, herein, experimental  LOC for estimating the permanent of an HPSM. From the linear optical experiments and theoretical analysis, we find that the LOC efficiency for a multiplicative error is dependent on the value of the permanent and that the lower bound of the computation time scales exponentially. Furthermore, our results are generalized and applied to LOC of permanents of unitary matrices, which can be implemented with a  multi-port  quantum interferometer involving single-photons at the input ports. We find that LOC with single-photons, for the permanent estimation, is on average less efficient than the most efficient classical algorithm known to date, even in ideal conditions. \end{abstract}

\maketitle


Quantum computers are expected to perform certain computational tasks exponentially faster than conventional digital computers \cite{Ladd10}.  Universal quantum computation relies on, along with other basic elements, quantum entanglement of a massive number of qubits. As preparation of large scale entanglement itself is already a daunting task, it is generally believed that universal quantum computers are unlikely to be available soon \cite{Preskill18}. Recent efforts are, thus, focused toward experimentally demonstrating the so-called ``quantum supremacy'' with a few tens of qubits  \cite{Lund17,Harrow17}. 

In photonic quantum information, quantum supremacy is often associated with BosonSampling which is the problem of generating samples following multi-mode interference among single-photon sources in a linear optical network~\cite{Aaronson13,Gard14,Gogolin13}. Since the sampling probability is related to the permanent of a complex matrix, a hard problem for classical digital computers \cite{Valiant79}, BosonSampling is often viewed as a linear optical pathway for achieving quantum supremacy  \cite{Spring13,Tillmann13,Spagnolo14,Bentivegna15,Wang17}. 

Recently, BosonSampling devices with input photon statistics other than single-photon states have been studied \cite{Rahimi15,Lund14,Hamilton17,Olson15,Rohde15}. In particular, it has been proposed that the permanent of a Hermitian positive semidefinite matrix (HPSM) can be estimated from a specific output probability of  thermal light through a linear optical network \cite{Chakh16,Rahimi15,Laibacher15}, although the exact calculation of the permanent  is known to be   a \#P-hard problem \cite{Grier16}. Linear optical computing (LOC) with thermal light therefore has been utilized in probing the computational complexity of the permanent of an HSPM. For instance, approximation of the permanent of an HPSM has been classified as a $\textrm{BPP}^{\textrm{NP}}$ problem using LOC  \cite{Rahimi15} and an LOC model was used to develop algorithm for  calculating permanents of a specific set of HPSMs \cite{Chakh16}.

In this work, we report an experimental  LOC for estimating the permanent of an HPSM. From the linear optical experiment with thermal light and theoretical analysis, we find that the LOC efficiency for a multiplicative error is dependent on the value of the permanent and  that the lower bound of the computation time scales exponentially with the size of an HPSM. Furthermore, our results are generalized and applied to LOC of the permanent of a unitary matrix, associated with a quantum interferometer involving a linear optical network and single-photon sources. We find that, even with ideal single-photon sources and without photon loss, linear optical computation for estimating the permanent of a unitary matrix is on average less efficient than the most efficient classical algorithm known to date. Although it already has been argued theoretically that linear optical computing of the permanent would require exponentially many samples \cite{Aaronson13,Gard14}, previous theoretical studies have been conceptually limited and could not give quantitative relationship for the computational cost on the matrix size and the value of permanent. In our work, from experimental and theoretical studies, we have provided a quantitative computational cost which can be compared against other algorithms. These results clearly demonstrate that linear optical computing of the permanent  does not offer advantages over classical approaches even when quantum resources are utilized.



Linear optical computing for estimating the permanent of an HPSM  involves the following three main components: an $M \times M$ linear optical interferometer $U$, thermal light at the $M$ input modes, and measurement of the coincidence detection probability at the $M$ output modes. The linear optical interferometer is built with a series of beam splitters and the thermal light state at the $i$-th input mode is described, in the Fock basis, by $\rho^{th}_{i} = \sum_{n=0}^{\infty}p^{th}_{i}(n)|n\rangle \langle n|$ where $p^{th}_{i}(n)=(1-\mu_{i})\mu_{i}^{n}$ and the mean photon number is $\langle n_{i}\rangle = \mu_{i}/(1-\mu_{i})$ with $0\leq\mu_{i}<1$. Finally, the probability that each of the $M$ output modes is occupied by a single photon is determined by measuring the $M$-fold joint detection probability  $p^{th}(1,1,...,1)$. Then, the permanent of a matrix $A$, Perm$[A]$, can be estimated from the joint detection probability $p^{th}(1,1,...,1)$ \cite{Chakh16, Rahimi15, Laibacher15},
\begin{subequations}
\begin{eqnarray}
&&\textrm{Perm}[A]= p^{th}(1,1,...,1)/\prod_{i=1}^{M}(1-\mu_{i}),\label{eq:01a}\\
&&A = UDU^{\dagger}, ~\textrm{where}~ D \equiv \text{diag}(\mu_{1},\mu_{2},...,\mu_{M}).\label{eq:01b}
\end{eqnarray}\label{eq:01}\end{subequations}
The matrix $A$ is an HPSM as the the eigenvalues, i.e. $\mu_{i}$, are greater than or equal to zero.

\begin{figure}[t]
\centering
\includegraphics[width=3.4in]{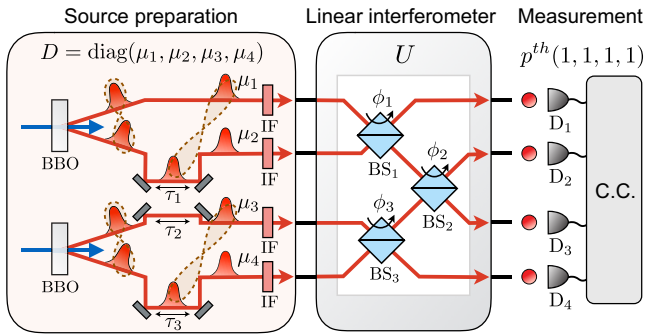}
\caption{Scheme for linear optical computing (LOC) with thermal light involving a $4\times4$ interferometer. The thermal sources having mean photon number of $\langle n_{i}\rangle = \mu_{i}/(1-\mu_{i})$ are filtered by 1-nm bandwidth interference filters (IF) and injected into the linear interferometer ($U$), composed of three beam splitters (BS$_i$). The linear optical transformation is determined by the transmissivities and the reflection phase shifts ($\phi_i$) of beam splitters. We measure the output distribution probability $p^{th}(1,1,1,1)$ from the coincidence count (C.C.) rate that all detectors (D$_j$) are clicked by a single photon. Temporal delays ($\tau_i$) are adjusted so that thermal pulses are overlapped at BSs. See main text for details.}\label{fig:01}
\end{figure}

\begin{figure}[t]
\centering
\includegraphics[width=2.8in]{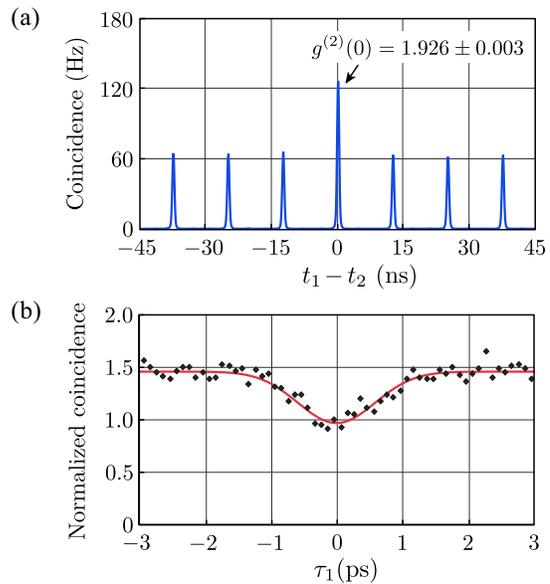}
\caption{Thermal pulse characterization and interference. (a) The  Hanbury-Brown-Twiss correlation measurement exhibits the thermal nature of the input pulses: $g^{(2)}(0) = 1.926 \pm 0.003$.  Here, $t_1 - t_2$ means the detection time difference between the output modes 1 and 2. (b) Two indistinguishable thermal pulses are overlapped at a 50:50 beam splitter (BS$_1$) by adjusting the time delay $\tau_1$. The time delay $\tau_1$ has an offset equal to the inverse repetition rate to choose two completely uncorrelated thermal pulses. The coincidence count, normalized by single counts of the two detectors, exhibits a dip when the two uncorrelated thermal pulses overlap at BS$_1$ and interfere. All data are acquired by using 1 nm full width at half maximum bandpass filters in front of the detectors.
}\label{fig:02}
\end{figure}

In the experiment, we have performed permanent estimation for various $2\times 2$ and $4\times 4$ HPSMs and the experimental setup is schematically shown in Fig.~\ref{fig:01}. An $M\times M$ unitary matrix $U$ is built with a series of beam splitters. Since the probability $p^{th}(1,1,...,1)$ needs to be determined experimentally, it is convenient that the experiment is repeated at a regular interval, requiring a pulsed source of thermal light. A pair of broadband  thermal light pulses can be prepared by using the spontaneous parametric down-conversion (SPDC) process. A frequency-doubled mode-locked Ti:Sa laser operating at the repetition rate of 80 MHz and at the pulse width of 140 fs is used as the pump. The pump laser has the central wavelength of 390 nm and 780 nm SPDC photon pairs are generated from a 2 mm thick type-I BBO crystal in the non-collinear geometry. Obviously, the SPDC photon pair is naturally energy-time entangled and is in a pure state. The state of each subsystem individually is described by partial trace over the unobserved system and therefore is found to be in a thermal state \cite{Yurke87,Strekalov99}. In experiment, we  choose two completely uncorrelated  thermal pulses by  introducing a relative optical delay equal to the inverse repetition rate, $\tau_1$ and $\tau_3$ in Fig.~\ref{fig:01}. The thermal pulses are first coupled to single-mode optical fibers and delivered to the input modes of the interferometer $U$. The thermal nature of the input state can be demonstrated by measuring the second-order correlation function $g^{(2)}(t_1-t_2)$  and the experimental data shown in Fig. \ref{fig:02}(a) clearly indicate the thermal nature of the input state with $g^{(2)}(0) = 1.926 \pm 0.003$ \cite{Hanbury56}.

\renewcommand{\arraystretch}{1.2}
\begin{table*}[tb]
\centering
\caption{Linear optical computing of the permanent of an HPSM. The unitary matrix ($U$) and the mean photon numbers at the input ($D$) are prepared according to the target HPSM $A=U D U^\dag$. The entries $U_{41}$, $U_{42}$, $U_{13}$, and $U_{14}$ are non-zero due to noise such as dark counts. $\text{Perm}[A]_{\textrm{exact}}$ refers to the exact mathematical calculation and $\text{Perm}[A]_{\textrm{exp}}$ refers linear optical computing with thermal light. Multi-mode interference is essential for linear optical computing and when the input thermal pulses are made to be distinguishable in time, incorrect permanent values are estimated as shown in ``No interference."  The  errors in $\text{Perm}[A]_{\textrm{exp}}$ are estimated from the statistical fluctuation of  photo-counts and refer to one standard deviation.}
\label{table:01}

\footnotesize 
\begin{tabular}{lllllllll}
\hline\hline  \\[-1.2em]
 \multicolumn{1}{c}{\small $U$} & & \multicolumn{1}{c}{\small $D\times10^3$} & & \multicolumn{1}{c}{\small $A \left(= UDU^{\dagger}\right)\times 10^3$}  & &\multicolumn{1}{c}{\small $\text{Perm}[A]_{\textrm{exact}}$} &   &\multicolumn{1}{c}{\small $\text{Perm}[A]_{\textrm{exp}}$} \\   \\[-1.2em] \noalign {\hrule height 0.7pt}   \\[-1em]
 
\multicolumn{1}{c}{\multirow{2}{*}{$\begin{pmatrix*}[r] 0.707 & 0.709\\ \minus 0.707&0.705 \end{pmatrix*}$}} & &
\multicolumn{1}{c}{\multirow{2}{*}{$\begin{pmatrix*}[r] 1.00 & 0.00\\ 0.00&1.04 \end{pmatrix*}$}} & &
\multicolumn{1}{c}{\multirow{2}{*}{$\begin{pmatrix*}[r] 1.02 & 0.02\\0.02&1.02\end{pmatrix*} $}} & &\multicolumn{1}{c}{$1.04 \times 10^{-6}$} & & \multicolumn{1}{c}{$(1.02\pm0.03) \times 10^{-6}$}     \\   \\[-1.1em]

\multicolumn{1}{c}{} & &
\multicolumn{1}{c}{} & &
\multicolumn{1}{c}{} & &\multicolumn{3}{c}{$(\text{No interference: } 1.56 \times 10^{-6})$}   \\   \\[-1.1em] \hline
 \\[-0.9em]
\multicolumn{1}{c}{\multirow{2}{*}{$\begin{pmatrix*}[r] 0.494 & 0.864\\ \minus 0.870&0.503 \end{pmatrix*}$}} & &
\multicolumn{1}{c}{\multirow{2}{*}{$\begin{pmatrix*}[r] 1.25 & 0.00\\ 0.00&1.92 \end{pmatrix*}$}}& &
\multicolumn{1}{c}{\multirow{2}{*}{$\begin{pmatrix*}[r] 1.74 & 0.30\\0.30&1.43 \end{pmatrix*} $}} & &\multicolumn{1}{c}{$2.58 \times 10^{-6}$} &  &
\multicolumn{1}{c}{$(2.54\pm0.04) \times 10^{-6}$} \\   \\[-1.1em] 

\multicolumn{1}{c}{} & &
\multicolumn{1}{c}{} & &
\multicolumn{1}{c}{} & &\multicolumn{3}{c}{$(\text{No interference: } 3.50 \times 10^{-6})$}   \\   \\[-1.1em] \hline
 \\[-2.2em]

\multicolumn{1}{c}{\multirow{2}{*}{$\begin{pmatrix*}[r] \minus 0.635&0.775&0.031&0.045\\ \minus 0.442&\minus 0.369&\minus 0.513&0.629 \\ 0.634 & 0.513& \minus 0.365 & 0.462\\0.021& 0.019&0.776&0.624\end{pmatrix*}$}}& &
\multicolumn{1}{c}{\multirow{2}{*}{$\begin{pmatrix*}[r] 1.66 & 0.00 & 0.00 & 0.00 \\ 0.00& 2.13&0.00&0.00 \\ 0.00 & 0.00 & 3.11 & 0.00 \\ 0.00 & 0.00 & 0.00 & 1.40 \end{pmatrix*}$}}  & &
\multicolumn{1}{c}{\multirow{2}{*}{$\begin{pmatrix*}[r] 1.95&\minus0.15&0.17&0.12\\ \minus0.15&1.99& 0.12 & \minus 0.72 \\0.17& 0.12&1.94&\minus 0.43\\ 0.12&\minus 0.72&\minus0.43&2.42\end{pmatrix*} $}} & & \multicolumn{1}{c}{$\begin{matrix*} \\ \\  21.4 \times 10^{-12}\end{matrix*}$}& & \multicolumn{1}{c}{$\begin{matrix*}\\ \\ (20.4\pm2.7) \times 10^{-12}\end{matrix*}$} \\   \\[-1.0em]

\multicolumn{1}{c}{} & &
\multicolumn{1}{c}{} & &
\multicolumn{1}{c}{} & &\multicolumn{3}{c}{$\text{(No interference: } 49.4 \times 10^{-12})$}   \\   \\[0 em] \hline
 \\[-2.2em]

\multicolumn{1}{c}{\multirow{2}{*}{$\begin{pmatrix*}[r] \minus 0.632&0.775&0.038&0.045\\ \minus 0.441& \minus 0.369&\minus0.517&0.629 \\ 0.636&0.513&\minus 0.359&0.462\\0.022&0.017&0.776&0.623\end{pmatrix*}$}}& &
\multicolumn{1}{c}{\multirow{2}{*}{$\begin{pmatrix*}[r] 1.57 & 0.00 & 0.00 & 0.00 \\ 0.00& 2.60&0.00&0.00 \\ 0.00 & 0.00 & 2.03 & 0.00 \\ 0.00 & 0.00 & 0.00 & 1.40 \end{pmatrix*}$}}  & & 
\multicolumn{1}{c}{\multirow{2}{*}{$\begin{pmatrix*}[r] 2.19&\minus 0.31& 0.40& 0.11\\ \minus 0.31&1.76&\minus 0.15&\minus 0.30 \\0.40&\minus 0.15&1.88&\minus 0.12\\ 0.11&\minus 0.30&\minus0.12&1.77\end{pmatrix*} $}} & &\multicolumn{1}{c}{$\begin{matrix*} \\ \\ 14.3 \times 10^{-12} \end{matrix*}$}& &\multicolumn{1}{c}{$\begin{matrix*} \\ \\ (14.8\pm2.3) \times 10^{-12}\end{matrix*}$} \\   \\[-1em] 

\multicolumn{1}{c}{} & &
\multicolumn{1}{c}{} & &
\multicolumn{1}{c}{} & &\multicolumn{3}{c}{$(\text{No interference: } 33.7 \times 10^{-12})$}   \\   \\[0 em] \hline\hline

\end{tabular}
\end{table*}

For Eq.~(\ref{eq:01}) to satisfy, all input thermal pulses to the linear optical network $U$ must be indistinguishable so that they interfere. Spectral indistinguishability is ensured by using 1 nm full width half maximum (FWHM) bandpass filters in front of the detectors. Temporally, relative optical delays ($\tau_1$, $\tau_2$, and $\tau_3$) are adjusted so that all input pulses are overlapped at the beam splitters. When two thermal states interfere at a beam splitter, the coincidence at the outputs of the beam splitter exhibits a dip as a function of the relative optical delay, similarly to Shih-Alley/Hong-Ou-Mandel interference, and this effect can be used to test spectral/temporal distinguishability between two input thermal states \cite{Shih86,Hong87,Chen11}. The experimental data for the two-photon interference with thermal light is shown in Fig.~\ref{fig:02}(b).  Two indistinguishable thermal pulses are overlapped at a 50:50 beam splitter (BS$_1$) by adjusting the time delay $\tau_1$. The coincidence  count, normalized by single counts of the two detectors,  exhibits a dip when the two uncorrelated thermal pulses overlap at BS$_1$ and interfere.  The ideal visibility in the thermal case is 1/3 and  the experimentally obtained visibility is $0.33\pm0.06$, indicating that the input thermal pulses are nearly identical \cite{Chen11}.


 For linear optical computing of the permanent of an HPSM $A= U D U^\dag$ with $D \equiv \text{diag}(\mu_{1},\mu_{2},...,\mu_{M})$,  we first need to prepare an $M\times M$ unitary matrix $U$ and thermal states with    $\langle n_i\rangle= \mu_{i}/(1-\mu_{i})$. In single-photon BosonSampling, photon loss is generally detrimental. However, in our case, the overall efficiency $\eta_i$ which includes channel loss and the detector efficiency, can be measured and pre-compensated by using the fact that  the mean photon number is linearly scaled,  i.e.  $\langle n_{i}\rangle \rightarrow \langle \eta_i n_i \rangle$. And,  according to Eq.~(\ref{eq:01}), accurate measurement of $p^{th}(1,1,...,1)$ crucial for estimating $\textrm{Perm}[A]$. Since we use non-number-resolving detectors, sufficiently weak thermal light pulses are used to suppress  multi-photon events contributing to the measurement of $p^{th}(1,1,...,1)$.  The similarity between the target HPSM $A$ and the experimentally prepared HPSM is checked by looking at the count rate of the detector  D$_j$. Slight intensity adjustment of the thermal pulses allows fine tuning of the experimental HPSM. As for the reflection phase shift $\phi_i$ from the beamsplitters, complete determination of all the phase values would not be trivial. However, the permanent of an HPSM is independent of the phases and therefore $\pi$ is assigned to all $\phi_i$. More experimental details can be found in the Appendices~\ref{sec:A}~and~\ref{sec:B}.
 
The experimentally implemented matrices $U$, $D$ and the resulting HPSM $A=UDU^\dag$ are shown in Table \ref{table:01}. To experimentally estimate the permanent, $\textrm{Perm}[A]$, according to Eq.~(\ref{eq:01}),  $p^{th}(1,1,...,1)$ is determined from the data accumulated for a period of 20 s and 36,500 s for $2\times2$ and $4\times4$ matrices, respectively. The experimentally estimated permanent values, obviously, have small errors which are  estimated from the statistical fluctuation of  photo-counts and refer to one standard deviation. The error depends on the number of samples and the value of the permanent. Nevertheless, linear optically computed values of the permanent agree quite well with the mathematical values, see Table \ref{table:01}. Note that multi-mode interference is essential for linear optical computing and, when the input thermal pulses are made to be distinguishable in time, incorrect permanent values are estimated as shown in ``No interference," see details in the Appendix~\ref{sec:C}.


The resource efficiency of linear optical computing for estimating the permanent of an HPSM can be studied by looking at the relationship between the total number of samples $N$, the dimension of the matrix $M$, and the output probability $p^{th}(1,1,...,1)$, the margin of error   $\varepsilon$, and the confidence level $\delta$ which is defined as \cite{Neyman37}   
 \begin{equation}
\delta = \textrm{Pr}[|\textrm{Perm}[A]_{\textrm{exp}}-\textrm{Perm}[A]_{\textrm{exact}}| < \varepsilon \textrm{Perm}[A]_{\textrm{exact}}],\nonumber
 \end{equation}
where $\textrm{Pr}[...]$ indicates the probability that the statement ``$...$" is true and  $ \varepsilon \textrm{Perm}[A]_{\textrm{exact}}$ represents the multiplicative error.

Consider now that, from Eq.~(\ref{eq:01}),  $\textrm{Perm}[A]$ is related to the probability that each detector clicks due to a single-photon $p^{th}(1,1,...,1)$. Thus the problem can be modeled as binomial sampling, which takes the value 0 or 1 with the probability   ``$1\minus p^{th}(1,1,...,1)$" or ``$p^{th}(1,1,...,1)$", respectively. Using the error bound of binomial sampling \cite{Brown01,Brown02}, the required number of samples $N$ to achieve the multiplicative error $\varepsilon \textrm{Perm}[A]$ is found to be, 
\begin{equation}
N=\frac{2 (\text{erf}^{-1}[\delta])^{2}(1-p^{th}(1,1,...,1))}{\varepsilon^2 p^{th}(1,1,...,1)},
\label{eq:02}
\end{equation}
where $\text{erf}^{-1}[x]$ is the Gauss inverse error function. The details can be found in the Appendix~\ref{sec:D}. At a glance, the total number of samples $N$ seems to be independent of the matrix dimension $M$. However, it turns out that the required number of samples scales as $\mathcal{O}(e^{M})$ because of the fact that $p^{th}(1,1,...,1) \leq e^{-M}$. For derivations, see the Appendix \ref{sec:E}.

\begin{figure}[t]
\centering
\includegraphics[width=3.1in]{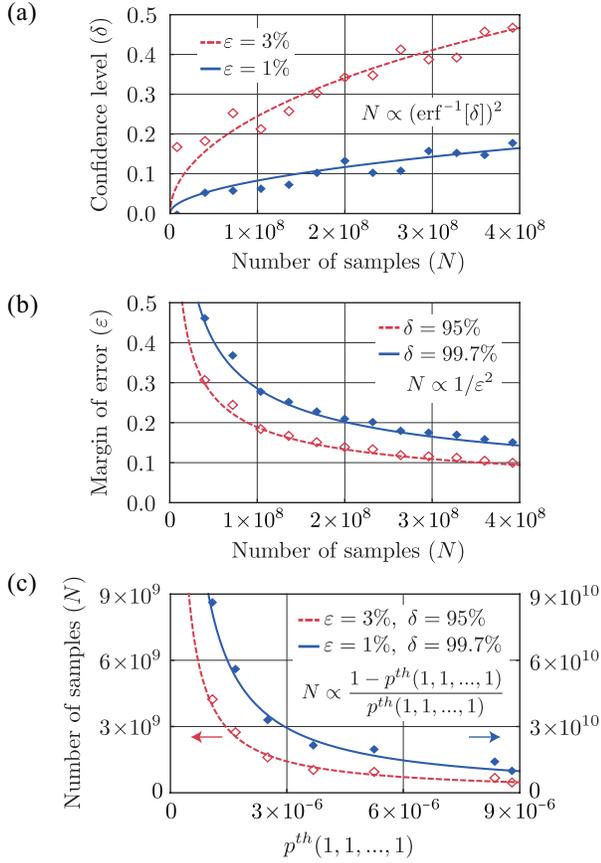} 
\caption{Efficiency and error analysis. For (a) and (b), diamond symbols represent the experimental data from the first row in Table \ref{table:01}. In (b),  $\delta$ values of 95\% and 99.7\% represent, respectively, two and three standard deviation confidence level in a Gaussian distribution. 
(c) The data points are obtained from the permanent estimation experiments for many different 2$\times$2 HPSMs. Note that the solid and dashed lines have different scales for the number of samples. All the lines are due to Eq.~(\ref{eq:02}).  
}
\label{fig:03}
\end{figure}

We find that the relation in Eq.~(\ref{eq:02}) agrees well with the experimental results. For instance, we applied Eq.~(\ref{eq:02}) to the data presented  in the first row of Table \ref{table:01} and the results are shown in Fig.~\ref{fig:03}(a) and \ref{fig:03}(b). Moreover, from the permanent estimation experiments for many different 2$\times$2 HPSMs, we obtain the relation between $N$ and  $p^{th}(1,1,...,1)$ as shown in Fig.~\ref{fig:03}(c). The experimental and theoretical results indicate that linear optical computing for estimating the permanent of an HPSM is on average inefficient.

The above analysis for the classical experiment (i.e., involving thermal states) can be applied to  the problem of  estimating the permanent of a unitary matrix $U$, which is implemented using a quantum interferometer with a linear optical network and single-photon sources. This is because the  probability that a single photon is detected in each output mode  $p^q(1,1,...,1)$ can still be modeled as binomial sampling with the only difference being $p^{q}(1, 1, ..., 1)=|\text{Perm}[U]|^2$ when a single-photon is injected in each input mode \cite{Aaronson13, Gard14}. Similarly to Eq.~(\ref{eq:02}), using the error bound of binomial sampling (see the Appendix~\ref{sec:D}) \cite{Brown01,Brown02},  the required number of samples $N$ to achieve the multiplicative error $\varepsilon |\textrm{Perm}[U]|^2$ is found to be, 
\begin{equation}
N=\frac{2 (\text{erf}^{-1}[\delta])^{2}(1-|\text{Perm}[U]|^2)}{\varepsilon^2 |\text{Perm}[U]|^2}.
\label{eq:03}\end{equation} 

Since the permanent of a unitary matrix is bounded as $|\text{Perm}[U]|\leq1$ regardless of the matrix dimension \cite{Aaronson14}, the lower bound of the number of required samples $N$ does not rise as the matrix dimension increases. At first sight, this result seems favorable. However, to meaningfully compare linear optical computing for estimating the permanent of a unitary matrix against classical algorithms, averaging over randomly selected unitary matrices is required. The averaged permanent $\langle|\text{Perm}[U]|^2\rangle_U$ in the entire unitary matrix space can be obtained from the random matrix theory and is given by \cite{Fyodorov,Drummond16}, 
\begin{equation}
\langle|\text{Perm}[U]|^2\rangle_U=\frac{(M-1)!M!}{(2M-1)!},
\end{equation}
where $M$ is the dimension of the unitary matrix.  As the matrix dimension $M$ increases, $\langle|\text{Perm}[U]|^2\rangle_U$ approaches to $\sqrt{4\pi M}/4^M$ asymptotically and, as a result, the required number of samples $N$ or the computation time for the multiplicative error scales as $\mathcal{O}(4^M/\sqrt{M})$. If we further consider the channel efficiency $\eta$,  the required number of samples $N$ increases by the factor $1/\eta^M$ which is necessary to reach the same level of statistical error as the coincidence probability is reduced. Comparing this result to that of the classical algorithm for exact permanent calculation which scales as  $\mathcal{O}(M^2 2^M)$ \cite{Aaronson14,Glynn10}, we find that quantum optical estimation of the matrix permanent, even in the ideal condition ($\eta=1$), is on average less efficient than  computation on a digital computer.



In summary, we have reported an experimental  linear optical computing for estimating the permanent of a HPSM with a linear optical network, thermal light, and single-photon counting measurement. We have shown that the error bound for linear optical computing of the permanent of a $M\times M$ HPSM is associated with the $M$-photon cross-correlation coincidence probability at the $M$ output modes. We have also shown that the lower bound of the linear optical computation time scales exponentially with $M$, demonstrating that linear optical computing is inefficient for estimating the permanent of a HPSM with a multiplicative error. Furthermore, we have found  that the error bound for linear optical computing of the permanent of a unitary matrix, associated with a quantum interferometer involving a linear optical network and single-photon sources, is also tied to the $M$-photon cross-correlation coincidence probability at $M$ output modes. This result indicates clearly that, even with ideal single-photon sources and without photon loss, linear optical computation for estimating the  permanent of a unitary matrix is on average less efficient than the most efficient classical algorithm known to date.

Although we have seen a lot of progress during the past decades, quantum computing is really in its infancy and we strongly believe that it is extremely important to theoretically and experimentally rule out applications that do not give advantages over classical approaches even though quantum resources are used. In our paper, we report a very thorough study on one such application, linear optical computing of the permanent based on multi-mode interference, and demonstrate that, even with single photon sources, the performance will not be better than classical approaches.\\


\section*{Acknowledgments}
This work was supported by the National Research Foundation of Korea  (Grants No. 2016R1A2A1A05005202, No. 2016R1A4A1008978 and No. 2015R1A6A3A04059773) Y. K. acknowledges the support from the Global Ph.D. Fellowship by the National Research Foundation of Korea (Grant No. 2015H1A2A1033028). 

\appendix
\section{Experimental estimation of the matrix permanent} \label{sec:A}
Assuming that  the optical channels are lossless and the detectors are on/off detectors with unity efficiency, the count rate at the detector $j$ due to the input thermal pulse at mode $i$ is given by 
\begin{equation}
C_{ij} = ft(p_{ij}(1)+p_{ij}(2)+p_{ij}(3)+...)=ft\mu_{ij},
\label{eq:05}
\end{equation}
where $f$ is the repetition rate and $t$ is the accumulation time. The average photon number at the input mode $i$ is $\langle n_i\rangle= \mu_{i}/(1-\mu_{i})$ and, after the interferometer, the resulting average photon number at the detector $j$ is $\langle n_{ij} \rangle= \mu_{ij}/(1-\mu_{ij})$.  The $n$-photon detection probability of detector $j$ due to the input thermal pulse at mode $i$ is $p_{ij}(n)=(1-\mu_{ij})\mu_{ij}^{n}$.  The relations $\langle n_i \rangle = \sum_j \langle n_{ij} \rangle$ holds as  the system is lossless and $C_i=\sum_j C_{ij}=ft \mu_i$ because $\sum_j \mu_{ij} \approx \mu_i$ under the condition $\mu_{i} \ll 1$. The matrix elements of the unitary matrix $U$ can then be determined from the detector counts as $U_{ji}= e^{i\varphi_{ji}}\sqrt{C_{ij}/C_{i}}$ from the unitarity condition $\sum_{j}{|U_{ji}|^2}=1$.  The phase factors $\varphi_{ji}$ cannot be measured here but, as they do not affect the permanent value, they are arbitrarily assigned to satisfy the unitarity condition.


The coincidence detection rate  when all output modes are populated with thermal light pulses is given by
\begin{equation}
C_{c}  = ft(p(1,1,...,1)+p(1,2,1,...)  + p(1,1,2,...)+...).
\label{eq:06}
\end{equation}
Under the condition  $\mu_{i} \ll 1$, multi-photon events are sufficiently suppressed and the above equation is approximated to $C_{c}  = ftp (1,1,...,1)$.

The permanent of a $M\times M$ Hermitian positive semidefinite matrix (HPSM) $A=UDU^{\dagger}$ can then be calculated from the experimentally measured values as
\begin{equation}
\textrm{Perm}[A]=C_c/\prod^M_{i=1}(ft-C_i).
\label{eq:07}
\end{equation}

\section{Freedom of phase assignment for unitary matrices} \label{sec:B}
There are degrees of freedom to assign the phases of unitary matrix $U$ without change of the permanent value, $\text{Perm}[A]$. That is, the matrix elements of $U$ can vary with the following relationship,
\begin{equation}
U_{ij}\rightarrow e^{i(\alpha_{i}+\beta_j)}U_{ij},
\label{eq:08}
\end{equation}
where $U_{ij}$ is an element of $U$ on the $i$-th row and $j$-th column, and $\alpha_{i}$ and $\beta_{j}$ are arbitrary phases.  The phase relation is obtained by introducing diagonal phase matrices $V$ and $W$ of $\text{diag}(e^{i\alpha_{1}},e^{i\alpha_{2}},...,e^{i\alpha_{M}})$ and $\text{diag}(e^{i\beta_{1}},e^{i\beta_{2}},...,e^{i\beta_{M}})$ for the unitary transformation $U\rightarrow VUW$. 

It is easy to see that $A=UDU^{\dagger}$ is invariant under $U \rightarrow UW$ transformation due to $WW^{\dagger}=\mathbb{I}$ and $W^{\dagger}DW=D$,  
\begin{align}
A&=U(WW^{\dagger})D(WW^{\dagger})U^{\dagger}=(UW)D(UW)^{\dagger}.
\label{eq:09}
\end{align}
Although the $U \rightarrow VU$ transformation changes $A$ to $A'=VAV^{\dagger}$, the value of $\textrm{Perm}[A']$ is the same with $\textrm{Perm}[A]$. The relation between $A'$ and $A$ is summarized as $A'_{ij}=A_{ij}e^{\text{i}(\beta_i-\beta_j)}$. Based on the relationship, the $\textrm{Perm}[A']$ can be written as, 
\begin{equation}
\textrm{Perm}[A']= \sum^{M!}_{\sigma \in S_n}\prod^{M}_{i=1} A_{i,\sigma(i)} e^{\text{i}(\beta_i-\beta_{\sigma(i)})},
\label{eq:10}
\end{equation}
where $\sigma(i)$ of the symmetric group $S_n$ is the $i$-th permutation of the set \{1, 2, ..., M\}. Since the permutation just changes the order of elements, $\sum_{i} \beta_{i}$ and $\sum_{i} \beta_{\sigma(i)}$ have the same value, i.e. $\prod_{i=1}^M e^{\text{i}(\beta_i-\beta_{\sigma(i)})}=1$. Accordingly, we can show that
\begin{equation}
\textrm{Perm}[A']=\textrm{Perm}[A]= \sum^{M!}_{\sigma \in S_n}\prod^{M}_{i=1} A_{i,\sigma(i)}.
\label{eq:11}
\end{equation}
Considering the above results, the permanent value is invariant under $U\rightarrow VUW$ and there are degrees of freedom for phase of $\alpha_i$ and $\beta_j$ for computing of the permanent of an HPSM.

To demonstrate linear optical computing of the permanent of an HPSM, the $4\times4$ unitary matrices were experimentally built with a series of beam splitter as shown in Fig. 1. The phase shift due to reflection is denoted by $\phi_i$ for each beamsplitter (BS$_i$) and the phase shift induced from the other side, if the BS is lossless, is $\pi-\phi_i$. Then, the $4\times4$ unitary matrix is written as 
\begin{equation} 
\small 
\begin{pmatrix} 
r_1 e^{i\phi_1}       & t_1                                             & 0       			                     & 0\\
t_1 r_2 e^{i\phi_2} & r_1 r_2 e^{i(\pi-\phi_1+\phi_2)} &r_3 t_2 e^{i\phi_3}			    &	t_3 t_2\\
t_1 t_2                   & r_1 t_2 e^{i(\pi-\phi_1)} 	       &r_3 r_2 e^{i(\pi-\phi_2+\phi_3)}  & t_3 r_2 e^{i(\pi-\phi_2)} \\
0                            & 0					       &t_3    					    & r_3 e^{i(\pi-\phi_3)}
\end{pmatrix}\nonumber
\end{equation}
where $t_{i}$ and $r_{i}$ are real numbers and the squares are the transmission and reflection ratios of BS$_i$, respectively. The ratios can be measured from the intensity splitting, but complete determination of all $\phi_i$ would not be trivial in experiment. Fortunately, the permanent of $UDU^{\dagger}$ is independent of $\phi_i$ and therefore $\pi$ is assigned to all $\phi_i$ in this work without loss of generality.


\section{Effect of temporal distinguishability} \label{sec:C}
The multimode interference is essential for linear optical computing and, when the input thermal pulses are made to be distinguishable in time, incorrect permanent values are estimated. In this section, we discuss the detection probability $p_{no}^{th}(1,1,...,1)$ that a single photon is detected in each output mode under ``No interference". For two-mode case, there are two possibilities contributing to $p_{no}^{th}(1,1)$; (i) two detectors are clicked due to the thermal pulses at a single mode, (ii) two detectors are clicked by the thermal pulses from each of the two modes. For the case (i), the bunching property of thermal light enhances the detection probability by a factor of two. However, the case (ii) doesn't have any enhancement in detection probability since there is no correlation between thermal pulses in different input modes. In summary, the total detection probability for two-mode under ``No interference" can be described as $p^{th}_{no}(1,1)=\sum^{2}_{i,j=1}e_{ij}|U_{1i}|^2|U_{2j}|^2 \mu_i \mu_j$ if $\mu_i\ll1$ where $U_{ji}$ is the element of a unitary matrix describing the transition from input mode $i$ to output mode $j$ and $\mu_i$ represents the mean photon number $\langle n_i \rangle=\mu_i/(1-\mu_i)$ of thermal light in input mode $i$. The $e_{ij}$ corresponds to the enhancement factor: $e_{ij}=2$ if $i= j$ and $e_{ij}=1$ if $i \neq j$.

Similarly, four-mode case has five situations; (i) four detectors are clicked due to the thermal pulses at a single mode, (ii) three detectors are clicked due to the thermal pulses at a single mode and another detector is clicked from the source of another input mode, (iii) two detectors are clicked from a single input mode source and the other two detectors are clicked from two different input modes, (iv) each pair of two detectors is clicked from the source of each of two input modes, (v) each detector is clicked from each different input mode. Enhancement factors $e_{ijkl}$ for four-mode case are 24, 6, 4,2 and 1 for (i-v), respectively~\cite{Stevens10}. By accumulating all contributions, the output probability is obtained as $p^{th}_{no}(1,1,1,1)=\sum^{4}_{i,j,k,l=1}e_{ijkl}|U_{1i}|^2|U_{2j}|^2|U_{3k}|^2|U_{4l}|^2 \mu_i \mu_j \mu_k \mu_l$ if $\mu_i\ll1$. 

To verify the multimode interference, the incorrect permanent values of $p^{th}_{no}(1,1,...,1)/\prod^M_{i=1}(1-\mu_i)$ are shown in Table 1.

\section{Resource efficiency of linear optical computing}  \label{sec:D}
The permanent of an HPSM can be estimated from the probability that each detector clicks due to a single-photon $p^{th}(1,1,...,1)$ with thermal light at input ports as shown in Eq. (\ref{eq:01}). The linear optical computing (LOC) with thermal light can be modeled as binomial sampling, which takes 0 or 1 with the probability of ``$1-p$" or ``$p$" where $p=p^{th}(1,1,...,1)$, respectively. If the number of samples $N$ is large enough, binomial distribution $B(N,p)$ can be approximated by the normal distribution  $\mathcal{N}(Np,\sqrt{Np(1-p)})$ where $Np$ and $\sqrt{Np(1-p)}$ are the mean value and standard deviation of the normal distribution, respectively~\cite{Brown01, Brown02}. 
The probability density function can be translated to standard normal distribution $\mathcal{N}(0,\sqrt{p(1-p)/N})$. It gives a confidence level as $\delta=\text{Pr}[|\hat{p}-p| <z_c\sqrt{p(1-p)/N}]$~\cite{Brown01,Brown02} where $\hat p$ is the estimated value by sampling and $z_c$ is the critical value for confidence level $\delta$ based on the standard normal distribution. Here, the term of $z_c\sqrt{p(1-p)/N)}$ means tolerable error and the critical value and confidence level are in the relationship of $z_c=\sqrt{2}\text{erf}^{-1}[\delta]$ where $\text{erf}^{-1}[x]$ is the Gauss inverse error function. For instance, confidence levels are 68\%, 95\% and 99.7\% when $z_c$ is 1, 2, and 3, respectively.


For a multiplicative error $\varepsilon p$, the margin of error $\varepsilon$ can be obtained by setting $\varepsilon p =z_c \sqrt{p(1-p)/N}$. Thus, the margin of error for LOC with thermal light is given as 
\begin{equation}
\varepsilon=\text{erf}^{-1}[\delta]\sqrt{\frac{2(1-p^{th}(1,1,...,1))}{N p^{th}(1,1,...,1)}}.
\label{eq:12}
\end{equation}
In another form, the required number of samples $N$ for a multiplicative error is
\begin{equation}
N=\frac{2 (\text{erf}^{-1}[\delta])^2 (1-p^{th}(1,1,...,1))}{\varepsilon^2 p^{th}(1,1,...,1)}.
\label{eq:13}
\end{equation}
At a glance, the total number fo samples $N$ seems to be independent of the matrix dimension $M$. However, it turns out that the required number of samples scales as $\mathcal{O}(e^M)$ because of the fact that $p^{th}(1,1,...,1)\leq e^{-M}$. See next section for details of the bound. In a strict sense, the above result is for the estimation of $p^{th}(1,1,...,1)$, but the required $N$ is the same for $\text{Perm}[A]$ because they have multiplicative relation of $p^{th}(1,1,...,1)=\textrm{Perm}[A]\times\prod_{i=1}^{M}(1-\mu_i)$.

If the largest eigenvalue of matrix $A$, $\mu_{max}$, is larger than 1, the matrix $A$ needs to be scaled as $A/\mu_{max}$ since $\mu_i>1$ is unphysical for LOC. For the case, $\text{Perm}[A]=\mu_{max}^M\text{Perm}[A/\mu_{max}]$ is estimated from the LOC of $\text{Perm}[A/\mu_{max}]$. Since this scaling also has multiplicative relation, the required $N$ does not change for the same margin of error and confidence level about a multiplicative error. 

If we consider an almost multiplicative error $\varepsilon \sqrt{p}$, scaling factors become relevant. If $\mu_{max}>1$ and the matrix $A$ is scaled as $A/\mu_{max}$, the confidence levels of $\text{Perm}[A]$ and $p^{th}(1,1,...,1)$ of $\text{Perm}[A/\mu_{max}]$ have the relationship of
\begin{widetext}
\begin{eqnarray}
\delta=&\text{Pr}&[|\text{Perm}[A]_{\text{samp}}-\text{Perm}[A]_{\text{exact}}|<\varepsilon \sqrt{\text{Perm}[A]}_{\text{exact}}]\nonumber\\
=&\text{Pr}&\left[|p^{th}(1,1,...,1)_{\text{samp}}-p^{th}(1,1,...,1)_{\text{exact}}|<\varepsilon \sqrt{\frac{p^{th}(1,1,...,1)_{\text{exact}}\prod^M_{i=1}(1-\mu_i/\mu_{max})}{\mu_{max}^M}}\right].
\label{eq:14}
\end{eqnarray}
\end{widetext}
As the result, the margin of error for almost multiplicative $\varepsilon \sqrt{\text{Perm}[A]}$ is found as
\begin{equation}
\varepsilon=\text{erf}^{-1}[\delta]\sqrt{\frac{2(1-p^{th}(1,1,...,1))\mu_{max}^M}{N\prod^M_{i=1}(1-\mu_i/\mu_{max})}},
\label{eq:15}
\end{equation}
and the required number of samples is given as
\begin{equation}
N=\frac{2 (\text{erf}^{-1}[\delta])^2 (1-p^{th}(1,1,...,1))\mu_{max}^M}{\varepsilon^2 \prod^M_{i=1}(1-\mu_i/\mu_{max})}.
\label{eq:16}
\end{equation}
It shows the required number of samples mainly depends on the $\mu_{max}$ and matrix dimension $M$ since $p^{th}(1,1,...,1)$ bounded by $e^{-M}$ does not significantly affect the required number of samples.

The above analysis for the classical experiment involving thermal light can be applied to the problem of estimating the permanent of a unitary matrix $U$, which is implemented using a quantum interferometer with a linear optical network and single-photon sources. This is because the probability that a single photon is detected in each output mode $p^{q}(1,1,...,1)$ can be modeled as binomial sampling with the only difference being $p^{q}(1,1,...,1)=|\text{Perm}[U]|^2$ when a single-photon is injected in each input mode~\cite{Aaronson13, Gard14}. Based on the analysis of LOC with thermal light, the required number of samples for permanent estimation of a unitary matrix about multiplicative error are given as
\begin{equation}
N=\frac{2 (\text{erf}^{-1}[\delta])^2 (1-|\text{Perm}[U]|^2)}{\varepsilon^2 |\text{Perm}[U]|^2}
\label{eq:17}
\end{equation}
by replacing $p^{th}(1,1,...,1)$ in Eq.~(\ref{eq:13}) with $p^{q}(1,1,...,1)=|\text{Perm}[U]|^2$. The scaling behavior about the matrix size are discussed in main text. 

Since LOC with single-photon sources does not require scaling factors, the required $N$ for almost multiplicative error are given as
\begin{equation}
N=\frac{2 (\text{erf}^{-1}[\delta])^2 (1-|\text{Perm}[U]|^2)}{\varepsilon^2}.
\label{eq:18}
\end{equation}
The $|\text{Perm}[U]|^2$ is less or equal to one for every matrix dimension, so that the required $N$ is bounded by $2(\text{erf}^{-1}[\delta]/\varepsilon)^2$ regardless of the matrix dimension.

\section{Upper and lower bounds of $\mathbf{p^{th}(1,1,...,1)}$}  \label{sec:E}
The required number of samples $N$, for LOC of the permanent of an HPSM, depends on the detection probability $p^{th}(1,1,...,1)$, see Eqns. (\ref{eq:13}) and (\ref{eq:16}). Thus we need to find out the bounds of $p^{th}(1,1,...,1)=\textrm{Perm}[A]\times\prod_{i=1}^{M}(1-\mu_i)$ to verify the scaling behavior of $N$ about the matrix dimension $M$. Here, $A=UDU^{\dagger}$ and $D=\text{diag}(\mu_1,\mu_2,...,\mu_M)$ where $0\le\mu_i<1$.  Considering the detection probability without input, the lower bound can be easily found as 0. However, the upper bound is nontrivial, so we try to find the upper bound based on physical intuition concerning the thermal light properties in linear interferometer. Two hypotheses are made to obtain the upper bound: (i) $p^{th}(1,1,...,1)$ will be maximized if the photons are uniformly distributed, (ii) a thermal light source is injected into only one input channel to avoid the photon bunching effect. The hypotheses give two conditions that $|U_{ij}|=1/\sqrt{M}$ and $\mu_i=0$ ($i\neq1)$, respectively. Under the conditions, $\text{Perm}[A]$ is calculated as $M!(\mu_1/M)^M$, accordingly $p^{th}(1,1,...,1)=M!(\mu_1/M)^M (1-\mu_1)$. The $p^{th}(1,1,...,1)$ is maximized with $\mu_1=M/(M+1)$, and it is given as
\begin{equation}
 \text{Max}[p^{th}(1,1,...,1)]=(\frac{1}{1+M})^{1+M}M!.
 \label{eq:19}
 \end{equation}
 
As the matrix dimension $M$ increases, the $p^{th}(1,1,...,1)$ asymptotically approaches $1/e^M$. Consequently, for a multiplicative error in Eq. (\ref{eq:13}), the lower bound of the number of samples $N$ or computation time scales exponentially and the computation time for an almost multiplicative error in Eq. (\ref{eq:16}) mainly depends on the scaling factors and matrix dimension.


\begin{thebibliography}{}

 \bibitem{Ladd10} T. D. Ladd, F. Jelezko, R. Laflamme, Y. Nakamura, C. Monroe, and J. L. O'Brien, Nature \textbf{464}, 45 (2010).

 \bibitem{Preskill18} J. Preskill, arXiv:1801.00862.

 \bibitem{Lund17} A. P. Lund, M. J. Bremner, and T. C. Ralph, npj Quantum Inf. \textbf{3}, 15 (2017).
 
 \bibitem{Harrow17} A. W. Harrow and A. Montanaro, Nature \textbf{549}, 203 (2017).
 
 \bibitem{Aaronson13} S. Aaronson and A. Arkhipov, Theory Comput. \textbf{9}, 143 (2013).
 
 \bibitem{Gard14} B. T. Gard, K. R. Motes, J. P. Olson, P. P. Rohde, and J. P. Dowling, arXiv:1406.6767.

 \bibitem{Gogolin13} C. Gogolin, M. Kliesch, L. Aolita, and J. Eisert, arXiv: 1306.3995.


 \bibitem{Valiant79} L. G. Valiant, Theoret. Comput. Sci. \textbf{8}, 189 (1979).

 \bibitem{Spring13} J. B. Spring, B. J. Metcalf, P. C. Humphreys, W. S. Kolthammer, X.-M. Jin, M. Barbieri, A. Datta, N. Thomas-Peter, N. K. Langford, D. Kundys, J. C. Gates, B. J. Smith, P. G. R. Smith, and I. A. Walmsley, Science \textbf{339}, 798 (2013).

 \bibitem{Tillmann13}  M. Tillmann, B. Daki\'c, R. Heilmann, S. Nolte, A. Szameit, and P. Walther, Nat. Photonics \textbf{7}, 540 (2013).
 
 \bibitem{Spagnolo14} N. Spagnolo, C. Vitelli, M. Bentivegna, D. J. Brod, A. Crespi, F. Flamini, S. Giacomini, G. Milani, R. Ramponi, P. Mataloni, R. Osellame, E. F. Galv\~ao, and F. Sciarrino, Nat. Photonics \textbf{8}, 615 (2014).
 
 \bibitem{Bentivegna15} M. Bentivegna, N. Spagnolo, C. Vitelli, F. Flamini, N. Viggianiello, L. Latmiral, P. Mataloni, D. J. Brod, E. F. Galv\~ao, A. Crespi, R. Ramponi, R. Osellame, and F. Sciarrino, Sci. Adv. \textbf{1}, e1400255 (2015).
 
 \bibitem{Wang17} H. Wang, Y. He, Y.-H. Li, Z.-E. Su, B. Li, H.-L. Huang, X. Ding, M.-C. Chen, C. Liu, J. Qin, J.-P. Li, Y.-M. He, C. Schneider, M. Kamp, C.-Z. Peng, S. H\"ofling, C.-Y. Lu, and J.-W. Pan, Nat. Photonics \textbf{11}, 361 (2017).

 \bibitem{Rohde15} P. P. Rohde, Phys. Rev. A \textbf{91}, 012307 (2015).

 
 \bibitem{Lund14} A. P. Lund, A. Laing, S. Rahimi-Keshari, T. Rudolph, J. L. O'Brien, and T. C. Ralph, Phys. Rev. Lett. \textbf{113}, 100502 (2014).
 
 \bibitem{Hamilton17} C. S. Hamilton, R. Kruse, L. Sansoni, S. Barkhofen, C. Silberhorn, and I. Jex, Phys. Rev. Lett. \textbf{119}, 170501 (2017).
 
 \bibitem{Olson15} J. P. Olson, K. P. Seshadreesan, K. R. Motes, P. P. Rohde, and J. P. Dowling, Phys. Rev. A \textbf{91}, 022317 (2015).
 

 \bibitem{Rahimi15} S. Rahimi-Keshari, A. P. Lund, and T. C. Ralph, Phys. Rev. Lett. \textbf{114}, 060501 (2015).


 \bibitem{Laibacher15} S. Laibacher and V. Tamma, Phys. Rev. Lett. \textbf{115}, 243605 (2015).
 
 \bibitem{Chakh16} L. Chakhmakhchyan, N. J. Cerf, and R. Garcia-Patron, Phys. Rev. A \textbf{96}, 022329 (2017).
 
 
 \bibitem{Grier16} D. Grier and L. Schaeffer, arXiv:1610.04670.



 \bibitem{Yurke87} B. Yurke and M. Potasek, Phys. Rev. A  \textbf{36}, 3464 (1987).

\bibitem{Strekalov99} D. V. Strekalov, Y.-H. Kim, and Y. Shih, Phys. Rev. A  \textbf{60}, 2685 (1999).

 \bibitem{Hanbury56} R. H. Brown and R. Q. Twiss, Nature \textbf{177}, 27 (1956).
 
\bibitem{Shih86} C. O. Alley and Y. H. Shih, in Proceedings of the Second International Symposium on Foundations of Quantum Mechanics in the Light of New Technology (Physical Society of Japan, Tokyo, 1986), pp. 47-52; Y. H. Shih and C. O. Alley, Phys. Rev. Lett. \textbf{61}, 2921 (1988).
 
 \bibitem{Hong87} C. K. Hong, Z. Y. Ou, and L. Mandel, Phys. Rev. Lett. \textbf{59}, 2044 (1987).
 
 \bibitem{Chen11} H. Chen, T. Peng, S. Karmakar, Z. Xie, and Y. Shih, Phys. Rev. A \textbf{84}, 033835 (2011).
 
 \bibitem{Neyman37} J. Neyman, Philos. Trans. Royal Soc. A \textbf{236}, 333 (1937).

 \bibitem{Brown01} L. D. Brown, T. T. Cai, and A. DasGupta, Stat. Sci. \textbf{16}, 101 (2001).

 \bibitem{Brown02} L. D. Brown, T. T. Cai, and A. DasGupta, Ann. Stat. \textbf{30}, 160 (2002).

 \bibitem{Aaronson14} S. Aaronson and T. Hance, Quantum Inf. Comput. \textbf{14}, 541 (2014).

 \bibitem{Fyodorov} Y. V. Fyodorov, Int. Math. Res. Notices \textbf{2006}, 61570 (2006).
 
 \bibitem{Drummond16} P. D. Drummond, B. Opanchuk, L. Rosales-Z{\'a}rate, M. D. Reid, and P. J. Forrester, Phys. Rev. A \textbf{94}, 042339 (2016).

 \bibitem{Glynn10} D. G. Glynn, Eur. J. Combin. \textbf{31}, 1887 (2010).
 
 \bibitem{Stevens10} M. J. Stevens, B. Baek, E. A. Dauler, A. J. Kerman, R. J. Molnar, S. A. Hamilton, K. K. Berggren, R. P. Mirin, and S. W. Nam, Opt. Express \textbf{18}, 1430 (2010).





 \end{thebibliography}
\end{document}